# IteRank: An iterative network-oriented approach to neighbor-based collaborative ranking

Bita Shams [a] and Saman Haratizadeh [a]
[a] *Faculty of New Sciences and Technologies, University of Tehran*
*North Kargar Street, Tehran, Iran. 14395-1374.*

## Abstract

Neighbor-based collaborative ranking (NCR) techniques follow three consecutive steps to recommend items to each target user: first they calculate the similarities among users, then they estimate concordance of pairwise preferences to the target user based on the calculated similarities. Finally, they use estimated pairwise preferences to infer the total ranking of items for the target user. This general approach faces some problems as the rank data is usually sparse as users usually have compared only a few pairs of items and consequently, the similarities among users is calculated based on limited information and is not accurate enough for inferring true values of preference concordance and can lead to an invalid ranking of items.

This article presents a novel framework, called IteRank, that models the data as a bipartite network containing users and pairwise preferences. It then simultaneously refines users' similarities and preferences' concordances using a random walk method on this graph structure. It uses the information in this first step in another network structure for simultaneously adjusting the concordances of preferences and rankings of items. Using this approach, IteRank can overcome some existing problems caused by the sparsity of the data. Experimental results show that IteRank improves the performance of recommendation compared to the state of the art NCR techniques that use the traditional NCR framework for recommendation.

## 1. Introduction

Recommender systems are the class of intelligent tools that exploit users' historical feedbacks in order to learn users' preferences and help them to find relevant contents. To do this, traditional recommender systems seek to generate a model that predicts users' ratings with minimal numerical error (e.g. RMSE). However, this approach does not guarantee accurate recommendation[1–5]. For example, consider a case that a target user $u$ would rate the movies "A" and "B" with 3 and 4 respectively, that is $r_A^u=3$ and $r_B^u=4$. Suppose that some recommendation algorithm predicts these ratings in the form of $r_A^u=4$ and $r_B^u=3$, while another algorithm predicts them in the form of $r_A^u=1$ and $r_B^u=2$. Although the second algorithm is less successful to reduce the rating prediction error, it will probably make a better recommendation since it correctly ranks "B" higher than "A".

Recently, a new generation of recommender systems, called Collaborative Ranking, has been emerged that seeks to minimize ranking prediction error instead of rating prediction error[2,6].

Collaborative ranking algorithms are categorized into two groups: latent factor collaborative ranking and neighbor-based collaborative ranking. Latent factor collaborative ranking (LCR) class of methods, exploits matrix factorization techniques to represent users and items in a latent feature space such that a ranking objective function be optimized[2,3,7–9]. On the other hand, neighbor-based collaborative ranking (NCR) methods estimate the ranking of the target user based on pairwise preferences of the users that are similar to him [1,4,6,10]. Although it is widely accepted that neighbor-based collaborative filtering is more applicable in commercial applications[11,12], NCR techniques has gained little attention in the research community.

This paper seeks to propose a novel framework that significantly improves the performance of the current neighbor-based collaborative ranking methods. Let $U$ be the set of users and $I$ be the set of items, NCR frameworks define the set of users' actions by a three-dimensional tensor $C_{|U|\times|I|\times|I|}$ such that

$$C_{u,i,j} = \begin{cases} 1, & u\ prefers\ item\ i\ over\ j \\ -1, & u\ prefers\ item\ j\ over\ i \\ 0, & Unknown \end{cases} \quad (1)$$

The matrix slice of $C$ corresponding to user $u$, represented by $C_u$, indicates the concordance matrix of $u$ that represents the concordance of the pairwise preferences with user $u$. To make a good recommendation to the target user $u$, it is necessary to design an approach for accurately estimating the unknown elements of $C_u$, and then, inferring the total rank of the unseen items based on the estimated values of $C_u$.

For this purpose, NCR algorithms generally follow a framework that is comprised of three modules running sequentially (See Fig.1).

1. *User-user similarity calculation:* These algorithms first calculate the similarity among the target user $u$ and other users based on Kendall correlation of common non-zero elements in *concordance matrices*.
2. *Concordance estimation:* In this step, the preferences of the similar users are used to estimate the concordance of each pairwise preference with the target user. More clearly, the *"concordance estimation"* module estimates the $C_{u,i,j}$ by calculating $\frac{\sum_{v\in Nei(u)} S_v^u C_{v,i,j}}{\sum_{v\in Nei(u)} |S_v^u|}$ where $Nei(u)$ is the neighbor set of $u$ and $S_v^u$ is similarity between $u$ and $v$.
3. In this final step, the calculated concordances for the preferences are used to infer a ranking for items using a greedy or random-walk approach.

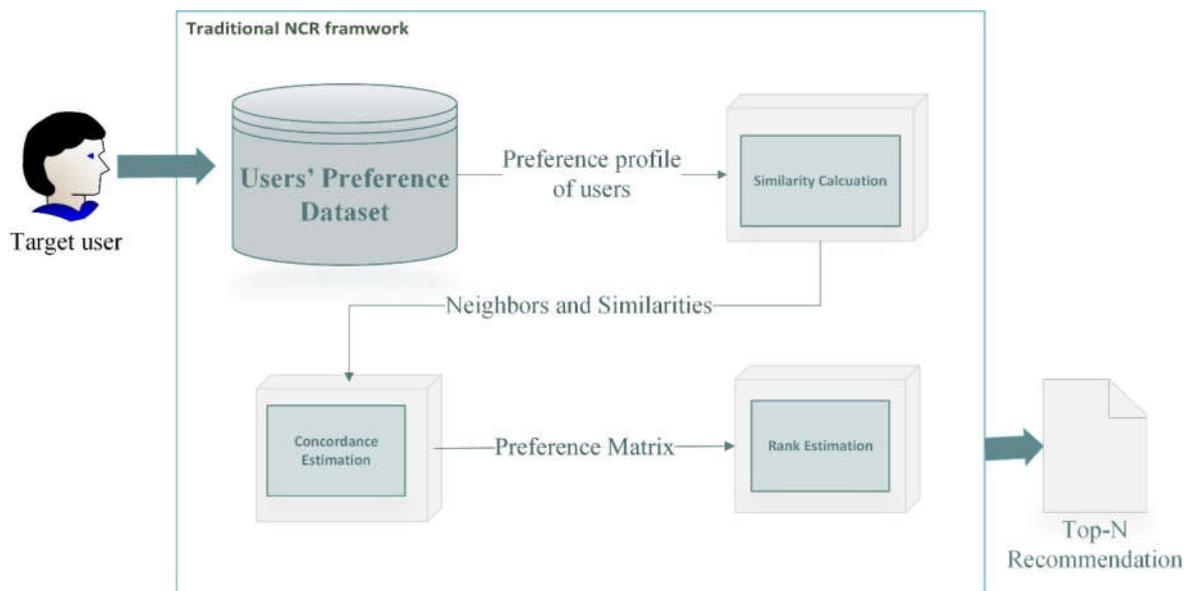

Figure 1. General framework for neighbor-based collaborative ranking

The performance of NCR framework in recommendation highly depends on how user-user similarities are calculated and how concordance estimations is done based on those calculated similarities. As explained before, in this traditional framework, this process is carried out in two separated steps, and that can lead to some shortcomings in the quality of calculated concordances.

- *Rare common preferences problem:*
  Most of the current NCR algorithms calculate users' similarities based on their direct agreements and disagreements over their common pairwise comparisons[1,6,13]. Unfortunately, in sparse data sets, users have few known preferences, and so they rarely have any pairwise comparisons in common. [4]. So, relying only on the users' common pairwise preferences in such data sets, extremely limits the power of similarity approximation process among users in two ways. First, some information is lost since a subset of preferences that are over other pairwise comparisons are totally ignored. Second, such an approach can find a small subset of users whose similarities to the target user is more than zero and can't differentiate among users who do not have any common pairwise comparisons with the target user[4].

- *Low discrimination flaw:*
  The second module in a typical NCR algorithm, uses the known preferences of the similar users to estimate the unknown preferences for the target user. However, even in larger neighborhoods, the known preferences for all the similar users still cover a small subset of all possible pairwise comparisons. That means that the concordance estimation module will not have a clue about the possible preference of a user about the most of pairwise comparisons. Hence, the *concordance estimation* method has to predict zero concordance for most of comparisons. Fig.2 illustrates that in average, such a typical NCR algorithm, has to report zero concordance for more than 96% of pairwise preferences and each target user. We refer to this fact as *low discrimination flaw* that is possibly one of the most important problems of NCR techniques.

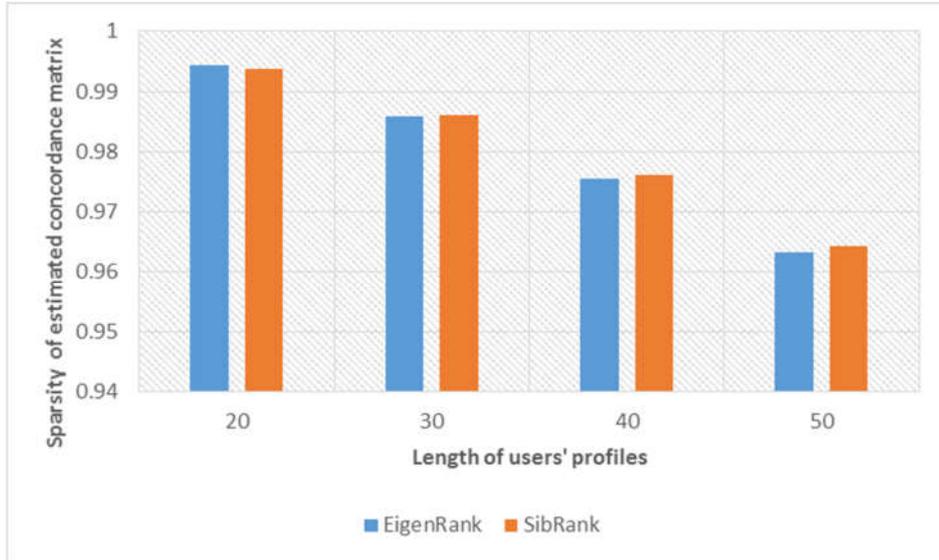

Figure 2. Average sparsity of preference matrix vs. length of users' profiles (i.e. number of rating given by each user) in Movielens100k dataset

This paper, contributes to present a novel framework, called IteRank that enables the *"concordance estimation"* module to iteratively communicate with other modules to adjust the concordance values using information it receives from them, and to propagate its own information to other modules for improving their performance as well. This flow of information, enables the system to solve the *rare common preferences problem* by expanding the profiles of the users based on the estimated concordances. Although using expanded profiles can be useful for reducing the effect of low discrimination flaw, IteRank also uses information from overall scores of the items for further refinement of concordance estimates so that it can more accurately estimate the target user's preferences over a larger subset of pairwise comparisons.

The key questions here is that how each module uses available information from other modules to improve itself. IteRank answers these questions by introducing two novel network structures and describing the behavior of a random surfer on those networks: It first models the preference data as a user-preference network and uses it to iteratively recalculate the users' similarities and preferences' concordances. Then, it exploits another network structure to simultaneously refine the concordance of preferences and the rank of the items. The resulting rank of the items can then be used to recommend the top N items to the user.

The main contributions of this paper can be summarized as below:

- We propose a novel iterative network-oriented approach for neighbor-based collaborative ranking. To our knowledge, this is the first algorithm that provides an ability to refine the similarity of users, concordance of pairwise preference, and rank of items from the target user's viewpoint in a level-wise iterative fashion.
- We introduce a novel network representation, called UPNet, for pairwise preference data sets that enables IteRank to simultaneously update the similarity of users and concordance of pairwise preferences.
- We also present a novel network representation called PRNet for modeling the relation among the pairwise comparisons and items. We use this network structure to iteratively aggregate the estimated pairwise preferences and infer the ranking of the items for the target user. Despite current approaches that uses a homogenous graph of items for ranking inference [4,6], IteRank exploits a bipartite preference-representative nodes for rank inference.

## 2. Related Work

Recommender systems use different types of information to improve the quality of personalized recommendation. Collaborative filtering takes into account the implicit/explicit feedbacks of users to items (e.g. rating, buys, clicks, etc.)[14] while content-based filtering considers the characteristics of users and items[15]. Trust-aware recommendation[16–18] benefits the social relationships among users of the system (e.g. trust, and distrust), and context-aware recommendation [19,20] uses contextual information such as time, location, and emotion at recommendation time. Though sidelong information (e.g. content, social relations, contextual) is useful for improving recommendation quality, it is not always available [21] and that is the reason why collaborative filtering is more popular than all other recommendation algorithms.

Collaborative filtering algorithms can be categorized into two groups: rating-oriented and ranking-oriented collaborative filtering. Rating-oriented algorithms emphasize on the absolute values of ratings, while, ranking-oriented algorithms, that are also called collaborative ranking methods, seek to learn the relative order of items [5]. It has been acknowledged that ranking oriented algorithms are more reliable for two reasons: First, recommendation algorithms requires to find which items are more relevant to users, no matter what their absolute rating values are. Second, relative order of items (i.e. ranking) may be simply obtained through tracing the users' interactions with systems. For instance, when a user clicks on *"iPhone "* as well as *"Samsung Note 7"and then* he buys *"Samsung Note 7"*, one can infer that the user has preferred *"Samsung Note 7" over "iPhone 7"*.

This paper presents a novel collaborative ranking algorithm that improves recommendation's accuracy. Here, we will briefly discuss the two streamlines of collaborative ranking techniques: matrix factorization (MFCR) and neighbor-based (NCR) algorithms.

### 2.1. Matrix factorization collaborative ranking

Matrix factorization collaborative ranking (MFCR) learns the latent representation of users and items that optimize a ranking-oriented loss function. MFCR algorithms generally follow one of two main objectives: They either focus on learning the ranking of relevant items or the pairwise comparisons between relevant and irrelevant items. Usually it is supposed that an item is relevant to a user *u* if u has had an interaction with it, and it is irrelevant if he has not.

The first approach assumes that recommender systems should accurately learn the ranking of relevant items. For instance, items with higher rating should be ranked higher than items with lower ratings. These algorithms, that are sometimes called NDCG-based MFCR algorithms, typically optimize the normalized discount cumulative gain of recommendation list when users have informed their priorities in terms of ratings or a preference list [3,5,21]. Cofi-Rank [2] is the pioneer algorithm in this category that optimizes a convex upper bound of NDCG. She et al.[3] seek to learn the latent factors that estimates top-1 probabilities of items in order to improve the ranking quality. URM unifies rating and ranking-oriented collaborative filtering by optimizing a convex combination of rating and ranking-oriented loss functions [5]. BoostMF [21] is a recent approach in this category that learns a set of matrix factorization models, each one optimizing an information retrieval metric over all rated items of users.

The second approach assumes that all relevant items are preferred over all irrelevant items, and so, a recommendation algorithm should learn to rank relevant items over irrelevant ones for each user. A subset of these algorithms, that are referred to as Bayesian personalized ranking (BPR) algorithms, learn the latent representation that is concordant to pairwise preferences among relevant and irrelevant items of users[7] and some researchers have used heterogeneous types of implicit feedback in this framework [22]. Beside BPR algorithms, there exist some other algorithms that optimize the mean reciprocal rank (MRR) of the recommendation list [8,23]. MRR reflects the average position of the first relevant item in the recommendation list. It should be noted that MRR-based algorithms emphasize on ranking of top items, while, the BPR algorithms take into account all pairwise preferences among relevant and irrelevant items. Recently, Christakopoulou et al. presented a novel approach that considers the pairwise comparisons of all items, while it weighs each pairwise comparison based on the rank of its corresponding items [24]. Note that unlike NDCG-based algorithms, this class of MFCR algorithms only differentiate between relevant and irrelevant items and neglect the difference between different types of interaction with a relevant item (i.e. click, buy)

We propose our algorithms at the intersection of BPR, and NDCG-based algorithms as it considers pairwise preferences while learning the ranking of relevant items. However the suggested algorithm is totally different from MFCR techniques as it does not use any matrix factorization approach. It lies in the category of neighbor-based collaborative ranking (NCR) algorithms as it exploits the opinions of similar users to predict the priorities of the target user.

2.2. Neighbor-based collaborative ranking

The traditional framework of neighbor-based collaborative ranking (NCR) is comprised of three modules: finding similar users, estimating the pairwise preferences of the target user, and inferring his total ranking. The majority of NCR techniques have tried to introduce accurate similarity measures. The most famous one is Kendall Correlation that measures the similarity the users based on their agreement and disagreement over pairwise comparisons[6]. EduRank[10] modified Kendall through defining a *compatible case*. The compatible case refers to the situation that one of the users have preferred both items equally. VSRank[1] improves Kendall correlation via considering the importance of each pairwise comparison in similarity calculation.

Once the similarities are calculated, NCR techniques estimate the concordance of each pairwise preference using opinions of the most similar users, called neighbors. Finally, they infer the total ranking of the items based on the estimated concordance values using a greedy[6,10,13,25] or a random walk approach [4,6,13].

The main challenge that this framework faces is the *sparsity* of the data used by recommender systems, that happens where users have only compared a small fraction of items and rarely have any pairwise comparisons in common. As we mentioned before, in such a situation, the approach of the traditional framework to calculating similarities and estimating concordance is not reliable enough and may lead to *Rare common preferences problem and low discrimination flaw*. Recently, some graph-based approaches have been proposed to cover these issue. SibRank [4] represents users' ranking as a signed bipartite preference network, called SiBreNet , and exploits it to calculate extended similarities even in case of no common pairwise comparison between the users. More clearly, SibRank somehow covers the *Rare common preferences problem* via considering indirect relations between users and extending the neighborhood. However, it still faces the *low discrimination flaw*. On the other hand, GRank [26] is another graph-based approach that seeks to directly rank the interest of the target user to all items *without emphasizing on similarity calculation and concordance estimation*. For this purpose, it constructs a heterogeneous graph consisting of three layers of users, pairwise preferences, and items' representatives. Then, it uses personalized PageRank to infer the ranking of target user over unseen items. Though GRank exhibited significant improvement over state-of-the art algorithms in dense data sets, its performance reduces in sparse data sets. That is because in GRank, the flow of information among different modules is not strictly controlled and when the number of unknown preferences increases in the heterogeneous network, they can mislead the personalized PageRank algorithm and cause inaccuracy in similarity calculation and final item ranking. Motivated by these approaches, this paper proposes another network-oriented framework, called IteRank, that is conceptually different from its ancestors, in the sense that it introduces and exploits two separate bipartite network structures for refinement of pairwise preference.

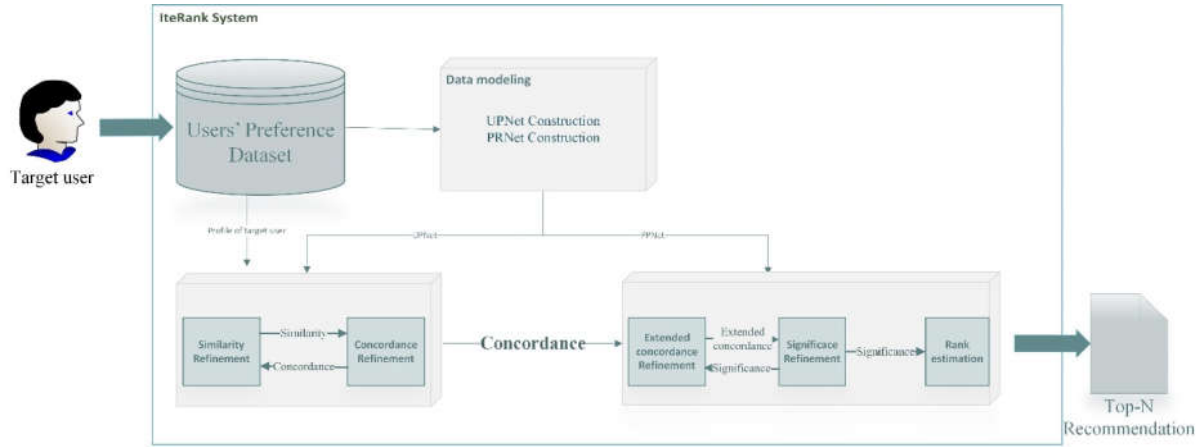

Figure 3. The framework of IteRank

## 3. IteRank

IteRank aims to accurately estimate the similarities, concordances, and rank of items. For this purpose, IteRank redesigns the NCR framework such that its modules can iteratively communicate each other in a forward and backward way (See Fig.3). This framework is comprised of two phases: at the first one, *"similarity refinement"* and *"concordance refinement"* modules communicate each other to adjust the similarity of users and concordance of pairwise preferences. Once the similarities and concordances converge, the concordance values are propagated to the next phase *in which "extended concordance refinement" and "significance refinement"* modules communicate to iteratively refine these concordances as well as the scores of the items. Note that this framework empowers NCR techniques to iteratively amend their performance using new information obtained in other modules. In the following sections, we elaborate how each module updates its content using available information from the input resources. Table 1 briefly introduces the basic mathematical notations of IteRank.

Table.1. Basic Mathematical Notations

| Notations | Definitions |
|---|---|
| U | The set of users |
| I | Set of items |
| P | Set of pairwise preferences |
| $p = <i,j>$ | A pairwise preference of item j over item j |
| C | A three dimensional tensor representing the users preference dataset |
| $C_u$ | The concordance matrix of user u |
| $C_{u,i,j}$ | The initial concordance of preference $p = <i,j>$ to the user u to |
| $C'_u$ | The estimated concordance matrix of user $u$ |
| $S_u$ | The vector of users' similarities to user $u$ |
| $S_u^v$ | Similarity of user $v$ to user $u$ |
| $i_d$ | Desirable representative of item $i$ |
| $i_u$ | Undesirable representative of item $i$ |
| R | Set of all desirable and undesirable representatives |

## 3.1. Phase 1: Measuring users' similarities and preferences' concordances.

In this phase, IteRank seeks to solve the rare common preferences problem through iteratively expansion of the target users' profile based users' similarities, and then, adjustment of the similarities based on expanded profile. For this purpose, IteRank seeks to design a procedure for similarity and concordance refinement. More formally, it defines the similarity S as a function of the elements if the concordance matrices C (i.e. $S = f(C)$). It also defines the estimated concordance values as a function of the users' similarities (i.e. $C'_u = g(S, C)$. We can define $f$ and $g$ through the following principles:

- *A user v is similar to u if v agrees with the preferences that are concordant with u.*
- *A pairwise preference p is concordant to the target user u if it is exposed directly by u or his similar users*

Trivially, this definition of similarity (S) and concordance (C) reminds the circular definition of graph-based ranking algorithms (e.g. PageRank[27]). Inspired by these algorithms, IteRank first models the preference dataset as a graph-structure, called UPNet, and then exploits this structure to obtain the similarity and concordance values. Fig.4b depicts the UPNet graph built from the preference dataset in Fig.4a. Note that preference dataset can be generated from all kinds of users' feedback including ratings

*Definition.1. User-Preference Network (UPNet) is a bipartite Graph G(U,P,E), where*
- *U is the set of users*
- $P = \{(i,j) | i \in I, j \in I, i \neq j\}$ *is the set of all pairwise preferences*
- $E_{UP} = \{< u, p > | u \in U, p = (i,j) \in P, C_{u,i,j} = 1\}$ *is the set of edges between users and pairwise preferences .*

Once that the UPNet is constructed (As shown in Fig.4b), the similarity and concordance values can be obtained based on the behavior of a random walker surfing the UPNet. This random walker randomly starts his walk from one of the target user's (direct) preferences. Then, he follows a random edge to find a *similar user* that believes in that preference, and, increases his similarity to the target user. Next, he keeps surfing a random edge to find a new *concordant preference* and increase its concordance. At each node, the random surfer might return to and restart from a direct preference of the user with probability $\alpha$ to emphasize the concordance of the target user's preferences. This process iterates several times after which, the users' similarities and preferences' concordance can be determined according to the relative frequency that the random surfer has reached the corresponding nodes.

To capture the behavior of the random surfer, IteRank represents the UPNET as a user-preference matrix $L_{|U| \times |P|}$ and preference-user $M_{|P| \times |U|}$ matrix that are respectively defined through Eq.2 and Eq.3

$$L_{up} = \frac{A_{up}}{\sum_{k=1}^{|U|} A_{kp}} \tag{2}$$

$$M_{pu} = \frac{A_{pu}^T}{\sum_{k=1}^{|P|} A_{ku}^T} \tag{3}$$

where $A^T$ is the transposed of UPNet's adjacency matrix. We should mention that both $L$ and $M$ are column stochastic matrices, $L_{up}$ represents the probability that a random walker move from a preference node $p$ to a user node $u$, while, $M_{pu}$ represents the probability of moving from node u to node $p$. Using $L$ and $M$, we can calculate the vector of users' similarities (S) and the vector of estimated preferences' concordance (c') for the target user $u$ by iteratively applying Eq. 4 and 5

$$S_u = (1-\alpha)LC'_u \tag{4}$$
$$C'_u = (1-\alpha)MS_u + \alpha d \tag{5}$$

Where $\alpha$ is the personalization factor, usually set to 0.15, and $d$ is the initial concordance vector that plays the role of the personalization vector and is defined by Eq. 6

$$d(j) = \begin{cases} L_{uj}, & A_{uj} = 1 \\ 0, & otherwise \end{cases} \tag{6}$$

Convergence: It is ensured that $s$ and $c$ will converge. More clearly, Eq.4 and Eq.5 can be rewritten in form of $z = (1-\alpha)Xy + \alpha v$, where $z = \begin{bmatrix} S \\ C' \end{bmatrix}$, $X = \begin{bmatrix} 0 & L \\ M & 0 \end{bmatrix}$ and v=$\begin{bmatrix} 0 \\ D \end{bmatrix}$. Since $X$ is a column stochastic matrix, and, $||y||_1 = ||z||_1 = 1$, it can be proved that $z$ will converge to the largest right eigenvector of the matrix $(1-\alpha)X + \alpha y1$ where 1 is a row vector of all ones [28].

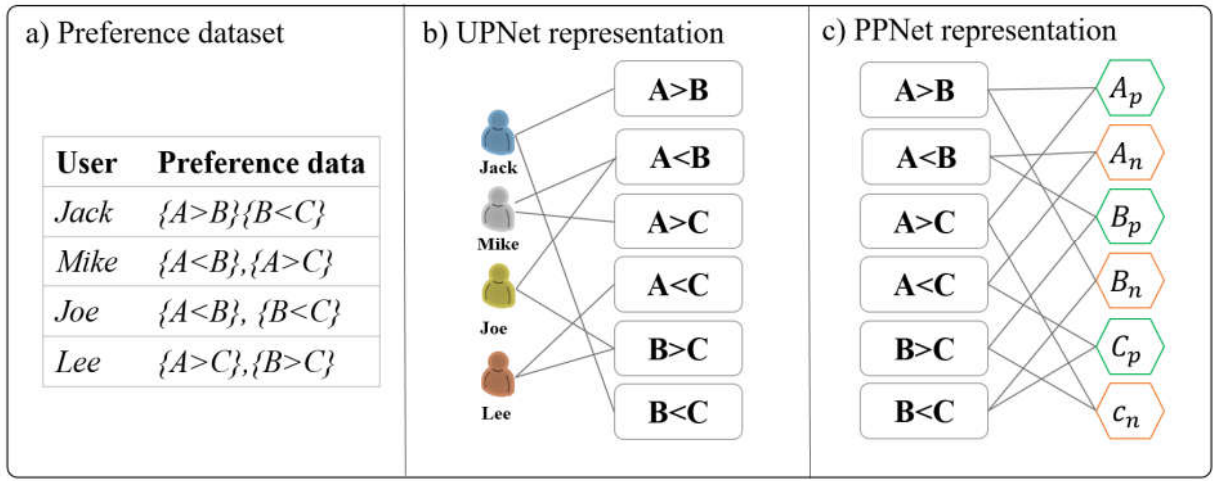

Figure 4. A schematic example to illustrate preference representation in IteRank

## 3.2. Phase2: Adjustment of preference concordance and inference of items' ranking

In this phase, IteRank uses information from overall scores of items to adjust the concordance values. In other words, IteRank lets concordance estimation and rank inference modules communicate to simultaneously refine the concordances of the preferences and scores of the items.

In the first phase of IteRank, some evidences are gathered to estimate the target user's preferences based on the known preferences of his neighbors. In this second phase, the gathered information is refined using evidences from overall desirability of items for that user: If item A is more desirable than item B based on the information gathered in the first phase, the system can conclude that the target user probably prefers $A$ over $B$, even if those two items have not been compared by the target user or any other users.

Generally, the desirability of an item $A$ is proportionate to the number of items over which $A$ is preferred and inversely proportionate to the number of items that are preferred over $A$. To model the desirability of items, IteRank exploits the concept of items' representatives [26]. Roughly speaking, the desirable representative of an item, $i_d$, is reinforced when the item wins at a comparison, while, its undesirable representative, $i_u$, is reinforced when it loses at a comparison. More formally, the relation between a preference $p = (a, b)$ and a representative $r$ is defined by the support function $sup: P \times R \to \{0,1\}$ such that $sup(p, r)$ indicates whether preference $p = (a, b)$ supports a representative $r$ or not.

$$sup(p,r) = \begin{cases} 1, & a = i \text{ and } r = i_d \\ 1, & b = i \text{ and } r = i_u \\ 0, & otherwise \end{cases} \qquad (7)$$

Following this intuition, IteRank adjusts the concordance of preferences to the target user while, at the same time, inferring the *significance* of items' representatives for him. The significance of each representative indicates how significantly the user believes in desirable or undesirable side of the item. To avoid ambiguity between concordance values obtained in different phases, we refer to the concordance values that are calculated in the second phase of IteRank, as extended concordance values.

Similar to the Phase1, IteRank presents a circular definition for extended concordance of a pairwise preference and significance of a representative as follows:

- *There exists an extended concordance between a pairwise preference and a target user, if that pairwise preference is concordant with the target user or supports representatives that are significant for the target user.*
- *An item's representative, desirable or undesirable, is significant for the target user, if there exists an extended concordance between the target user and pairwise preferences that support that representative.*

According to these definitions, IteRank constructs a graph structure, called PRNet that models the relations between preferences and items' representatives according to Definition 2.

*Definition.2. Preference-Representative Network (PRNet) is a bipartite Graph G(P,R, $E_{PR}$), where*

- $P = \{<i,j> | i \in I, j \in I, i \neq j\}$ is the set of all pairwise preferences
- $R = \{i_p | i \in i_d \cup i_u | i \in I\}$ denotes the set of desirable and undesirable representatives of items.
- $E_{PR} = \{(p,r) | sup(p,r) = 1, p \in P, r \in R\}$ is the set of edges that link the preference nodes to the representative nodes.

Fig.4c depicts a schematic example of PRNet. In the data set containing 3 items, there are $2 * \binom{3}{2}$ preference node and 6 representatives. Each preference node $\{A < B\}$ is connected to two representative nodes $A_n$ and $B_p$ since it supports the desirable representative of B and the undesirable representative of A.

Based on the structure of PRNet, we now describe how significance of the representatives and extended concordance of the preferences can be obtained through the behavior of a random walker. This random walker starts the walk from one of the concordant preferences in PRNet. Then, it follows a random edge to reach a representative node that is supported by that concordant preference and keeps surfing PRNet to find new preferences supporting that representative. To avoid fading the concordance vector calculated in UPNet, the random walker will occasionally return to one of the initial concordant preferences with probability $\beta$, while, the chance of restarting from any concordant preference will be proportional to its concordance value.

To model the random walker's behavior, IteRank first represents PRNet as a preference-representative matrix $W$ and representative-preference matrix $T$ through Eq.8 and Eq.9.

$$W_{pr} = \frac{B_{pr}}{\sum_{k=1}^{|P|} B_{kr}} \tag{8}$$

$$T_{rp} = \frac{B_{rp}^T}{\sum_{k=1}^{|R|} B_{kp}^T} \tag{9}$$

where $B_{|P| \times (|I|)}$ as the adjacency matrix of *PRNet*, that represents the relations between preferences and representatives in PRNet. Note that W is a column stochastic matrix and $W_{pr}$ indicates the probability that a random walker in representative node $r$ will move to pairwise preference node $p$. Similarly, $T_{rp}$ indicates the probability of moving from node $p$ to node $r$. After defining W and T, it calculates the vector of extended concordance values *(h)* and the vector of significance values*(p)* through Eq.10 and Eq.11

$$h = (1-\beta)Wp + \beta q \tag{10}$$
$$p = (1-\beta)Th \tag{11}$$

Where $\beta$, usually set to 0.15, is the personalization factor, and $q$ is the personalization vector that indicates initial concordance of preference. We define q though Eq. 12

$$q = \frac{c}{\sum_{i=1}^{|P|} c_i} \quad (12)$$

Where $c$ is the final concordance vector that is obtained through Eq. 3 and 4. Similar to the similarities/concordance vector, it can be shown that that $h$ and $p$ will converge too.

After significance of items' representatives are calculated, IteRank estimates the interest of the target user to the item $i$ through Eq. 13

$$score(i) = \frac{p(i_p)}{p(i_p) + p(i_N)} \quad (13)$$

## 3.3. Algorithm Summarization

Algorithm. 1 summarizes the IteRank's framework. IteRank first constructs the transition matrix corresponding to UPNet (Line 2-3) and initiates the similarity *(s)*, concordance *(c)*, and personalization *(d)* vectors (Line 4-6). Then, it iteratively updates *s* and *c* through Eq. 5 and Eq. 6 (Line 8-11). When *c* and *s* converge, IteRank starts to amend the concordance of preferences and infer the user's judgment on items' representative; for this purpose, it defines the transition matrix and the initial vector of extended concordance values (line 13-14). Then, it initiates (line 15-17) and updates the vectors of significance and the extended concordance values through Eq.9 and Eq. 10 (line 19-22). Finally, IteRank calculates items' scores using Eq. 13 and recommends the top-k items according to their scores (line 23-27).

Algorithm 1. IteRank's Pseudo-code

|    | Input: UPNet graph, PRNet graph, Target user u, number of recommended items K<br>Output: The best k items. |
|----|---|
| 1  | //Phase 1 |
| 2  | L, M: Initialize transition matrices of UPNet L, M based on Eq. 2 and Eq.3 |
| 3  | $d$: initialized personalization vector based on Eq.6 |
| 4  | S, C: random initialization of S and C |
| 5  | $S_0 = S_0/(S_0 + C_0)$ |
| 6  | $C_0 = C_0/(S_0 + C_0)$ |
| 7  | $t = 0$ |
| 8  | Until converge |
| 9  | $\quad s_{t+1} = (1 - \alpha)Lc_t$ |
| 10 | $\quad c_{t+1} = (1 - \alpha)Ms_t + \alpha d$ |
| 11 | $\quad t = t + 1$ |
| 12 | //Phase 2 |
| 13 | W,T :Initialize transition matrices of PRNet based on Eq. 8 and Eq.9 |

| 14 | $q = \dfrac{c_t}{\sum_{i=1}^{|P|} c_t(i)}$ |
|---|---|
| 15 | $p,h$: random initialization of $p$ and $h$ |
| 16 | $p_0 = p_0/(p_0+h)$ |
| 17 | $h_0 = h_0/(p_0+h_0)$ |
| 18 | $t' = 0$ |
| 19 | Until converges |
| 20 | $\quad h_{t'+1} = (1-\beta)Wp_{t'} + \beta q$ |
| 21 | $\quad p_{t'+1} = (1-\beta)Th_{t'}$ |
| 22 | $\quad t' = t' + 1$ |
| 23 | // Calculating items' relevance |
| 24 | For each item $i$, |
| 25 | $\quad score(i) = \dfrac{p_{t'}(i_d)}{p_{t'}(i_d) + p_{t'}(i_u)}$ |
| 26 | Sort items according to their score values discerningly in recommendation list RL |
| 27 | Return the best k items in RL |

## 3.4. Computational Complexity

Assume that $N_U$ is the number of users, $N_I$ is the number of items, and $N_P$ is the number of pairwise preferences assigned by all users. Clearly, the total number of preference nodes is equal to $e = N_I^2$ and the number of representative nodes is $2N_I$.

To calculate computational complexity of IteRank, let us take a deeper look at each phase. In the first phase, we update the similarity and concordance values of each node using the following equations:

$$S_v = (1-\alpha)\sum_{i \in Nei(v)} \frac{1}{|Nei(i)|} C'_i \qquad (14)$$

$$C'_i = (1-\alpha)\sum_{v \in Nei(i)} \frac{1}{|Nei(v)|} S_v + \alpha d(i) \qquad (15)$$

Where *Nei(x)* is the set of adjacent nodes to *x*. So, the value corresponding to each node $x$ is updatedat computational complexity of $|Nei(x)|$. Hence, the total number of operations in each iteration will be $\sum_{x \in UPNet} |Nei(x)| = 2 E_{UPNet}$. More precisely, each iteration is calculated at computational complexity of $O(N_P)$ since the number of UPNet's edges is $N_P$.

Similarly, the computational complexity of refining *extended concordance* and significance values is equal to $O(E_{PPNet}) = O(N_I^2)$. To sum up, the computational complexity of IteRank is $O(tN_P + t'N_I^2)$. As we know, users have explored a small set of items and consequently, have only compared a small set of total comparisons. Thus, $N_p$ will be equal to $cN_I^2$ for some small constant $c$ (e.g. 2.48 for some experiments on movielens100k). Therefore, the computational complexity of IteRank is equal to $O(t'N_I^2)$. It is worth noting that both *t'* and *t* do not exceed 20 in our experiments.

Note that graph construction does not significantly increase the computational complexity in practical applications since the structure of the networks are not related to the target user, and can be cached in the memory. Furthermore, these graphs can be updated efficiently in case of availability of new information, such as introducing a new user, a new item or a new pairwise preference:

- When an old user assigns a new pairwise preference, UPNet is updated at *O(1)* by inserting a new edge between the user and the pairwise preference. It is worth noting that most of the updates in the recommender systems lies in this category.
- When a new user arrives, UPNet is updated at *O(1)* by creating a new node.
- When a new item is inserted into the system, both UPNet and PRNet is updated at $O(N_I)$ by creating $2*N_I$ nodes corresponding to the pairwise comparison between the new item and other ones in addition to creating $2*(2*N_I)$ edges to link these nodes to the corresponding representative nodes.

The computational complexity of current NCR technique is in the order of $O(N_U N_I^2 + K N_I^2 + N_I^2)$. More clearly, the computational complexity is $O(N_U N_I^2)$ for similarity calculation, $O(K N_I^2)$ for concordance estimation, and $O(N_I^2)$ for rank inference [4]. So, IteRank would not increase the computational complexity compared to the current NCR techniques. (e.g. EigenRank, SibRank). The computational complexity of IteRank is also comparable to model-based algorithms that focus on learning users' pairwise preferences or the ranking of items. These algorithms [7,29] seek to minimize a loss function representing incorrect prediction of users' pairwise preferences. For this purpose, they require to iteratively examine all pairwise preferences of users at the computational complexity of $O(N^2)$. Therefore, their total computational complexity will be $O(tN^2)$ where *t* is the number of iterations till convergence [30]. Furthermore, updating latent factor models costs as much as its training, while, IteRank, can updates its graph structure with O(1) in case of inserting new preferences or user, and, O(N) in case of inserting new items. It should be noted that IteRank's running time can be significantly reduced through parallel implementation of sparse matrix multiplication in distributed frameworks. Therefore, IteRank is also applicable to real-world applications including large number of users and items.

Table.2. Statistical properties of datasets

| Dataset | #users | #items | $\#\frac{ratings}{user}$ | $\#\frac{ratings}{items}$ | Sparsity | Average of rating | Variance of rating |
|---|---|---|---|---|---|---|---|
| ML-100K | 943 | 1682 | 106.64 | 59.45 | 0.0630 | 3.52 | 1.26 |
| ML-1M | 6040 | 3952 | 165.59 | 253.08 | 0.0410 | 3.58 | 1.24 |
| Epinions | 5358 | 5000 | 13.43 | 14.39 | 0.0030 | 4.69 | 0.45 |
| MT | 47271 | 27061 | 12.18 | 21.26 | 0.0004 | 7.30 | 3.45 |

## 4. Experiments
### 4.1. Evaluation Framework

We have conducted a comprehensive set of experiments following an evaluation protocol widely used for collaborative ranking techniques([2–5,31]). For each user, a fixed number $UPL = \{10,20,30,40,50\}$ of ratings is randomly sampled and placed on the training set, and the remaining ratings are put in the test set. To assess algorithms based on their Top10 recommendations, we make sure that each user has at least 10 items in his test set. For each UPL value, we have repeated the whole experiments 5 times and reported the average and standard deviation. That wipes out the impact of randomness in sampling training set.

This evaluation protocol is widely used in collaborative ranking literature as evaluation is based on higher number of test items than training items (i.e. UPL). Therefore, it simulates the real world applications in which recommender systems should only suggest a small number of items among numerous unseen items for each user [24,31]. Furthermore, varying UPL provides an ability to analyze performance of algorithms under different situations regarding the number of users, the number of items, and the degree of sparsity of data.

#### 4.1.1. Datasets

We conducted all experiments on three publicly available datasets, Movielens, Epinions, and MovieTweeting that are widely used in related works [1,3–5]. Movielens datasets have been generated with different number of users, items, and, ratings. We also compare the algorithms over Epinions dataset originally generated by Massa and Avesani [18]. In our experiments, we filter this dataset by randomly sample 5000 items that are rated by more than 50 users. MovieTweetings (MT), is a movie rating dataset that is originally introduced in ACM RecSys 2014 challenge [32]. Here, we use the version, released on Feb, 2017, that contains IMDB ratings of 27,016 items which are shared by 47,142 users on Twitter. Unlike MovieLens and Epinions data in which the rates are mostly high, MovieTweetings is comprised of ratings scattered from 1 to 10. Table 2 describes the statistical properties of datasets. For all NCR techniques, the users' feedbacks in the datasets have been converted from ratings to pairwise comparisons.

Table 3. Comparison of algorithms in ML100k dataset

| Dataset | Algorithm | NDCG@1 | NDCG@3 | NDCG@5 | NDCG@10 |
|---|---|---|---|---|---|
| UPL=10 | IteRank | 0.611±0.023 | 0.614±0.030 | 0.623±0.02 | 0.651±0.017 |
| | SibRank | 0.630±0.004 | 0.621±0.010 | 0.622±0.01 | 0.650±0.009 |
| | GRank | 0.588±0.025 | 0.589±0.02 | 0.595±0.03 | 0.632±0.024 |
| | EigenRank | 0.575±0.025 | 0.569±0.020 | 0.573±0.03 | 0.601±0.024 |
| | CofiRank | 0.593±0.047 | 0.595±0.020 | 0.602±0.02 | 0.631±0.013 |
| UPL=20 | IteRank | 0.679±0.016 | 0.665±0.021 | 0.664±0.020 | 0.675±0.017 |
| | SibRank | 0.676±0.020 | 0.664±0.014 | 0.660±0.015 | 0.672±0.012 |
| | GRank | 0.650±0.016 | 0.641±0.021 | 0.642±0.022 | 0.657±0.020 |
| | EigenRank | 0.654±0.022 | 0.645±0.018 | 0.642±0.020 | 0.655±0.017 |
| | CofiRank | 0.587±0.008 | 0.598±0.011 | 0.603±0.014 | 0.620±0.024 |
| UPL=30 | IteRank | 0.721±0.009 | 0.708±0.008 | 0.706±0.008 | 0.712±0.005 |
| | SibRank | 0.704±0.011 | 0.691±0.010 | 0.686±0.010 | 0.692±0.010 |
| | GRank | 0.698±0.016 | 0.686±0.013 | 0.685±0.013 | 0.694±0.010 |
| | EigenRank | 0.696±0.007 | 0.682±0.010 | 0.679±0.010 | 0.686±0.005 |
| | CofiRank | 0.593±0.011 | 0.605±0.008 | 0.607±0.008 | 0.623±0.012 |
| UPL=40 | IteRank | 0.739±0.012 | 0.726±0.006 | 0.722±0.004 | 0.725±0.007 |
| | SibRank | 0.722±0.008 | 0.706±0.009 | 0.700±0.009 | 0.705±0.006 |
| | GRank | 0.736±0.007 | 0.717±0.005 | 0.710±0.005 | 0.715±0.005 |
| | EigenRank | 0.713±0.012 | 0.700±0.006 | 0.695±0.006 | 0.695±0.004 |
| | CofiRank | 0.595±0.017 | 0.606±0.009 | 0.607±0.006 | 0.619±0.012 |
| UPL=50 | IteRank | 0.746±0.005 | 0.730±0.005 | 0.723±0.004 | 0.722±0.005 |
| | SibRank | 0.731±0.013 | 0.719±0.004 | 0.711±0.006 | 0.71±0.004 |
| | GRank | 0.742±0.006 | 0.727±0.003 | 0.719±0.004 | 0.717±0.005 |
| | EigenRank | 0.708±0.015 | 0.701±0.006 | 0.697±0.006 | 0.697±0.005 |
| | CofiRank | 0.599±0.025 | 0.611±0.009 | 0.609±0.010 | 0.616±0.004 |

4.1.2. Baseline algorithms

To evaluate the performance of the proposed framework to the three-step NCR framework, we compare IteRank to EigenRank, SibRank and GRank, the state of the art NCR algorithms. For further analysis, we assess its performance against CofiRank, the most well-known latent factor model for collaborative ranking.

- EigenRank: EigenRank [6] is the most famous algorithm in the family of NCR techniques. We have implemented the random-walk version of EigenRank using neighborhood sizes of 10 and 100 and $\alpha = 0.85$ that are best reported parameters for EigenRank.
- SibRank: SibRank [4] is the state of the art NCR algorithm that represents the user's ranking as a signed bipartite network and uses signed multiplicative rank propagation to calculate users' similarities. We have implemented SibRank using neighborhood sizes 100 and $\alpha = 0.85$ corresponding to the best results reported for SibRank.
- GRank: GRank [26] is another state of the art algorithm that exploits personalized PageRank on a tripartite graph structure to directly estimate the users' interests to items. We have implemented GRank by setting $\alpha = 0.85$ as suggested in the original work.

- CofiRank: CofiRank [2] learns the latent representation for users and items by optimizing a convex upper bound of a loss based on Normalized Discounted Cumulative Gain (NDCG). We used the publicly available code for CofiRank and adopted the optimal values for its parameters as suggested in [33]

### 4.1.3. Evaluation Metric

We evaluate the collaborative ranking algorithms through measuring their Normalized Cumulative Discounted Gain (NDCG). NDCG compares cumulative relevance of items in the recommendation list and the ideal one. More specifically, the NDCG of Top-N recommendation list for user *u* is computed using Eq. 16

$$NDCG@TopN_u = \frac{1}{\delta_u} \sum_{i=1}^{topN} \frac{2^{rel_i^u} - 1}{\log(i+1)} \tag{16}$$

Where $rel_i^u$ is the relevance grade assigned by *u* to *i-th* items in the generated recommendation list, and, $\delta_u$ is the normalization factor that ensures the NDCG would be equal to 1 for the ideal recommendation list.

### 4.2. Performance Analysis

We assess the performance of the recommendation algorithms regarding the length of users' profiles and recommendation lists. As depicted in Table 3-6, IteRank outperforms other algorithms in the majority of evaluation conditions. Here we summarize and discuss the most important results of the experiments:

- Experimental results demonstrate superiority of IteRank over other NCR and graph-based algorithms. IteRank achieves NDCG@10 of 63% in case of UPL=30 in MovieTweetings dataset, while the performance of SibRank is 60%, and EigenRank and GRank achieve a 59% performance. Also, IteRank achieves performance of 95% in case of UPL=10 in Epinions datasets, while, the next best performance is 87% for GRank. Though SibRank reaches higher performance in terms of NDCG@1 when UPL is 10 in ML100K, and ML1M, its performance is noticeably lower than IteRank in higher UPL values and higher NDCG Levels.
- In most of the experiments, IteRank significantly outperforms CofiRank except in ML1M when UPL is equal to 10. In such a situation, CofiRank can achieve the highest performance because of the high degree of sparsity of the data as well as the low diversity of ratings in the training sets (See Table. 3). In such a situation, MFCR algorithms perform better than NCR techniques as they train a model that learns the overall interest of users. That also explains why CofiRank outperforms EigenRank and SibRank in Epinions dataset (See Table.5). However, the superiority of CofiRank vanishes when the data gets denser or the rates are more diverse. As shown in Table.4, CofiRank performs up to 14% worse than

IteRank, 13% worse than GRank, 10% worse than SibRank, and 4% worse than EigenRank in case of UPL=50 in ML1M. A similar trend can be observed in MovieTweetings datasets where ratings are more diverse than other datasets. In this dataset, IteRank and other NCR algorithms exhibit higher performance compared to CofiRank in all UPL values (See Table. 6).

Table 4. Comparison of algorithms in ML1M dataset

| Dataset | Algorithm | NDCG@1 | NDCG@3 | NDCG@5 | NDCG@10 |
|---|---|---|---|---|---|
| UPL=10 | IteRank | 0.653±0.015 | 0.651±0.017 | 0.654±0.016 | 0.670±0.015 |
|  | SibRank | 0.681±0.006 | 0.671±0.007 | 0.67±0.009 | 0.675±0.009 |
|  | GRank | 0.637±0.021 | 0.636±0.023 | 0.638±0.022 | 0.654±0.019 |
|  | EigenRank | 0.621±0.004 | 0.608±0.005 | 0.605±0.005 | 0.613±0.004 |
|  | CofiRank | 0.708±0.002 | 0.69±0.006 | 0.683±0.003 | 0.685±0.001 |
| UPL=20 | IteRank | 0.732±0.007 | 0.718±0.011 | 0.711±0.0011 | 0.710±0.0011 |
|  | SibRank | 0.717±0.013 | 0.706±0.008 | 0.701±0.007 | 0.701±0.006 |
|  | GRank | 0.708±0.013 | 0.696±0.013 | 0.692±0.013 | 0.694±0.014 |
|  | EigenRank | 0.711±0.022 | 0.706±0.000 | 0.700±0.001 | 0.699±0.001 |
|  | CofiRank | 0.690±0.010 | 0.679±0.008 | 0.676±0.009 | 0.685±0.024 |
| UPL=30 | IteRank | 0.774±0.010 | 0.759±0.008 | 0.750±0.007 | 0.746±0.006 |
|  | SibRank | 0.732±0.024 | 0.720±0.015 | 0.714±0.013 | 0.711±0.011 |
|  | GRank | 0.752±0.005 | 0.737±0.005 | 0.730±0.004 | 0.728±0.002 |
|  | EigenRank | 0.720±0.016 | 0.709±0.006 | 0.704±0.008 | 0.703±0.009 |
|  | CofiRank | 0.680±0.012 | 0.676±0.014 | 0.669±0.012 | 0.670±0.016 |
| UPL=40 | IteRank | 0.789±0.015 | 0.770±0.013 | 0.762±0.013 | 0.756±0.012 |
|  | SibRank | 0.737±0.020 | 0.726±0.018 | 0.719±0.016 | 0.716±0.014 |
|  | GRank | 0.771±0.015 | 0.756±0.013 | 0.748±0.012 | 0.746±0.010 |
|  | EigenRank | 0.719±0.018 | 0.706±0.010 | 0.701±0.012 | 0.701±0.012 |
|  | CofiRank | 0.662±0.008 | 0.661±0.006 | 0.656±0.005 | 0.657±0.003 |
| UPL=50 | IteRank | 0.788±0.028 | 0.772±0.018 | 0.764±0.017 | 0.758±0.015 |
|  | SibRank | 0.749±0.020 | 0.734±0.017 | 0.727±0.016 | 0.723±0.013 |
|  | GRank | 0.778±0.021 | 0.763±0.020 | 0.757±0.019 | 0.755±0.014 |
|  | EigenRank | 0.697±0.020 | 0.698±0.009 | 0.693±0.010 | 0.692±0.012 |
|  | CofiRank | 0.647±0.005 | 0.645±0.005 | 0.642±0.005 | 0.645±0.003 |

Now we will generally discuss how varying the size of users' profiles will affect the performance of IteRank compared to other algorithms. Every trend in the results that is discussed here, is general and can be observed over all NDCG values.

Experimental results demonstrate that higher values of UPL lead to higher improvement of IteRank over CofiRank in all datasets. As an example, IteRank performs about 2%, 10%, 12%, better than CofiRank in terms of NDCG@ 5 when UPL=10, 30, and 50 in ML100K, respectively. This is a fairly expected result, as MFCR algorithms learn a model that generally fits into the whole set of the training data and so can easily capture the overall desirability of the items. Hence, MFCR algorithms perform efficiently in sparse data where users have compared a small group of items and the information is only enough to recognize the popular items that are usually in top of the users' favorite lists. On the other hand, in denser datasets, the local and personal tastes of the users are more distinguishable and accordingly, IteRank and other NCR techniques that investigate the local neighborhoods perform better than the MFCR algorithms.

Table 5. Comparison of algorithms in Epinions dataset

| Dataset | Algorithm | NDCG@1 | NDCG@3 | NDCG@5 | NDCG@10 |
|---|---|---|---|---|---|
| UPL=10 | IteRank | 0.92±0.008 | 0.936±0.010 | 0.943±0.011 | 0.950±0.010 |
| | SibRank | 0.775±0.008 | 0.803±0.005 | 0.814±0.005 | 0.838±0.004 |
| | GRank | 0.827±0.006 | 0.842±0.006 | 0.845±0.004 | 0.856±0.002 |
| | EigenRank | 0.779±0.007 | 0.805±0.003 | 0.814±0.002 | 0.837±0.001 |
| | CofiRank | 0.850±0.021 | 0.861±0.024 | 0.867±0.024 | 0.879±0.022 |
| UPL=20 | IteRank | 0.923±0.016 | 0.936±0.014 | 0.943±0.013 | 0.953±0.010 |
| | SibRank | 0.759±0.015 | 0.798±0.014 | 0.813±0.014 | 0.842±0.010 |
| | GRank | 0.858±0.009 | 0.872±0.007 | 0.878±0.005 | 0.888±0.003 |
| | EigenRank | 0.775±0.005 | 0.804±0.004 | 0.816±0.004 | 0.839±0.004 |
| | CofiRank | 0.900±0.027 | 0.912±0.026 | 0.913±0.024 | 0.916±0.019 |
| UPL=30 | IteRank | 0.911±0.014 | 0.928±0.013 | 0.937±0.012 | 0.948±0.010 |
| | SibRank | 0.737±0.014 | 0.783±0.012 | 0.808±0.011 | 0.842±0.009 |
| | GRank | 0.895±0.004 | 0.904±0.003 | 0.908±0.004 | 0.913±0.003 |
| | EigenRank | 0.749±0.010 | 0.789±0.006 | 0.807±0.006 | 0.834±0.006 |
| | CofiRank | 0.892±0.020 | 0.906±0.019 | 0.909±0.019 | 0.911±0.017 |
| UPL=40 | IteRank | 0.911±0.015 | 0.929±0.012 | 0.937±0.010 | 0.948±0.009 |
| | SibRank | 0.754±0.027 | 0.789±0.017 | 0.813±0.014 | 0.845±0.010 |
| | GRank | 0.907±0.005 | 0.919±0.002 | 0.924±0.002 | 0.929±0.002 |
| | EigenRank | 0.737±0.026 | 0.776±0.018 | 0.798±0.014 | 0.830±0.010 |
| | CofiRank | 0.880±0.024 | 0.891±0.020 | 0.894±0.021 | 0.894±0.014 |
| UPL=50 | IteRank | 0.903±0.015 | 0.925±0.015 | 0.935±0.012 | 0.947±0.011 |
| | SibRank | 0.728±0.026 | 0.781±0.021 | 0.808±0.016 | 0.845±0.011 |
| | GRank | 0.906±0.007 | 0.922±0.004 | 0.928±0.002 | 0.934±0.003 |
| | EigenRank | 0.715±0.037 | 0.766±0.029 | 0.792±0.021 | 0.827±0.014 |
| | CofiRank | 0.881±0.033 | 0.894±0.021 | 0.892±0.019 | 0.896±0.018 |

Also, we can observe that increasing UPL will also enhance improvement of IteRank over SibRank and EigenRank in ML100K, ML1M, and Epinions. For instance, in case of UPL=10, both of IteRank and SibRank achieves equal performance (i.e. 67%) in terms of NDCG@10, while, IteRank exhibits 3% improvement over SibRank when UPL is equal to 50. On the other hand, IteRank gains higher improvement in lower UPL values in MovieTweetings dataset. That observation can be explained through special characteristics of these datasets. In Movielense and Epinions datasets, the rates are high and they have a low variance. In such datasets, most of users have similar opinions about items and increasing UPL will provide valuable information about the overall interest of items. Therefore, IteRank would enhance its performance in two ways: first, it uses the farther available information to estimate an overall ranking of items. Second, its benefits its refinement procedure to adjust the similarity, concordance, and significance values. However, the story is different in MovieTweetings data set where ratings are more diverse. In such a dataset, using overall information will be beneficial in case of sparse data where information is not enough to learn the diversity among users' ratings. Unlike other datasets, increasing the length of profiles will not reflect the overall interest of items. Instead, it would reflect the diversity among local tastes of users that is beneficial to NCR algorithms. Therefore, improvement of IteRank will be reduced in higher UPL values in MovieTweetings datasets. For instance, in case of UPL=10, IteRank exhibits an improvement of 6% over EigenRank in terms of NDCG@5, while, its improvement will be about 3% when UPL is 50.

Table 6. Comparison of algorithms in MovieTweetings dataset

| Dataset | Algorithm | NDCG@1 | NDCG@3 | NDCG@5 | NDCG@10 |
|---|---|---|---|---|---|
| UPL=10 | IteRank | 0.497±0.007 | 0.534±0.004 | 0.566±0.000 | 0.625±0.002 |
| | SibRank | 0.496±0.011 | 0.511±0.010 | 0.531±0.005 | 0.584±0.002 |
| | GRank | 0.487±0.010 | 0.489±0.024 | 0.505±0.013 | 0.556±0.007 |
| | EigenRank | 0.450±0.010 | 0.462±0.003 | 0.481±0.000 | 0.537±0.001 |
| | CofiRank | 0.460±0.010 | 0.488±0.003 | 0.515±0.002 | 0.576±0.003 |
| UPL=20 | IteRank | 0.530±0.010 | 0.545±0.002 | 0.571±0.002 | 0.627±0.002 |
| | SibRank | 0.523±0.013 | 0.535±0.019 | 0.553±0.001 | 0.600±0.000 |
| | GRank | 0.510±0.011 | 0.521±0.018 | 0.537±0.004 | 0.581±0.007 |
| | EigenRank | 0.504±0.017 | 0.519±0.000 | 0.538±0.000 | 0.583±0.000 |
| | CofiRank | 0.455±0.023 | 0.478±0.004 | 0.505±0.002 | 0.562±0.001 |
| UPL=30 | IteRank | 0.534±0.011 | 0.553±0.001 | 0.579±0.004 | 0.632±0.003 |
| | SibRank | 0.530±0.012 | 0.546±0.008 | 0.563±0.004 | 0.607±0.004 |
| | GRank | 0.538±0.012 | 0.545±0.008 | 0.556±0.009 | 0.596±0.006 |
| | EigenRank | 0.523±0.012 | 0.537±0.014 | 0.553±0.008 | 0.596±0.004 |
| | CofiRank | 0.450±0.013 | 0.469±0.004 | 0.494±0.002 | 0.551±0.006 |
| UPL=40 | IteRank | 0.514±0.009 | 0.553±0.010 | 0.580±0.006 | 0.627±0.003 |
| | SibRank | 0.534±0.012 | 0.545±0.024 | 0.563±0.002 | 0.603±0.000 |
| | GRank | 0.535±0.039 | 0.536±0.014 | 0.542±0.039 | 0.578±0.036 |
| | EigenRank | 0.525±0.008 | 0.54±0.000 | 0.558±0.000 | 0.600±0.001 |
| | CofiRank | 0.423±0.057 | 0.449±0.007 | 0.463±0.002 | 0.500±0.000 |
| UPL=50 | IteRank | 0.524±0.011 | 0.562±0.011 | 0.582±0.007 | 0.627±0.003 |
| | SibRank | 0.540±0.014 | 0.555±0.003 | 0.572±0.004 | 0.611±0.003 |
| | GRank | 0.529±0.013 | 0.536±0.023 | 0.55±0.015 | 0.594±0.007 |
| | EigenRank | 0.525±0.018 | 0.539±0.009 | 0.555±0.003 | 0.594±0.003 |
| | CofiRank | 0.435±0.002 | 0.456±0.017 | 0.48±0.005 | 0.533±0.004 |

Performance analysis of IteRank and GRank shows that IteRank achieves higher improvement over GRank in lower UPL values. As an example, in Epinions, IteRank shows an improvement of 7% over GRank in terms of NDCG@1 when UPL is 10, while, it is only 2% better in case of UPL=50 (See Table.5). This trend relies on the structure of tripartite preference graph (TPG) that GRank uses for recommendation. As the length of users' profiles decreases, the structure of TPG gets more unbalanced. Since GRank does not explicitly manage the flow of information among different layers of TPG, this skewness affects the performance of GRank and makes its final recommendation less reliable.

## 4.3. Time Analysis

We also analyze the running time and scalability of IteRank compared to other algorithms. As running time depends on several factors, we first investigate the computational complexity of an up-to-date recommendation in real world application. EigenRank, as a traditional NCR algorithms, does not require any model construction. Though SibRank, GRank, and IteRank construct a graph, they can simply keep the graph up-to-date through inserting each new instance of data (See Section 3.**). Therefore, all NCR algorithms will make an up-to-date recommendation, and their overall computational complexity is mainly defined by the approach they use to calculate the interest of the target user to each item. Unlike NCR algorithms, model-based algorithms require to reconstruct the model to keep it up-to-date. Therefore, updating the model may cost as much as the initial model construction in model-based algorithms such as CofiRank.

Accordingly, running time of an up-to-date recommendation in model-based recommendation depends on their model construction, while, that of a NCR algorithm is related to their recommendation algorithms. As model construction is too expensive, model-based algorithms periodically reconstruct the model which may result in loss of new information in some intervals. Due to this fundamental difference, it is not plausible to compare the running time of these classes of algorithms [35]. As this paper focuses on NCR algorithms, we compare the running time of involved NCR algorithms.

Here, we report the running time for a simple implementation of IteRank to compare its speed with other NCR algorithms. The execution times are measured in seconds on a Linux based PC running an Intel core i7-5820K processor at 3.3GHz with 32GB of RAM. IteRank makes recommendation to the target user in less than a second in ML100K which is up to 30 times less than the running time of SibRank. Please note that model-based collaborative ranking algorithms requires at least several minutes to construct a model in this dataset. As shown in Fig.5, IteRank's recommendation gets much faster than other NCR algorithms as UPL increases. The reason is that in higher values of UPL, more information is available from the neighbors of each user, and collecting and aggregating that information will increase the running time of the traditional NCR algorithms (i.e. SibRank and EigenRank). On the other hand, IteRank's running time is quite stable as its computational complexity mainly depends on the number of items in the dataset that is not significantly affected by the UPL value.

We reemphasize that using GPU programming for matrix multiplication phases (Lines 9-10, 20-21) can speed up IteRank more than 10 times.

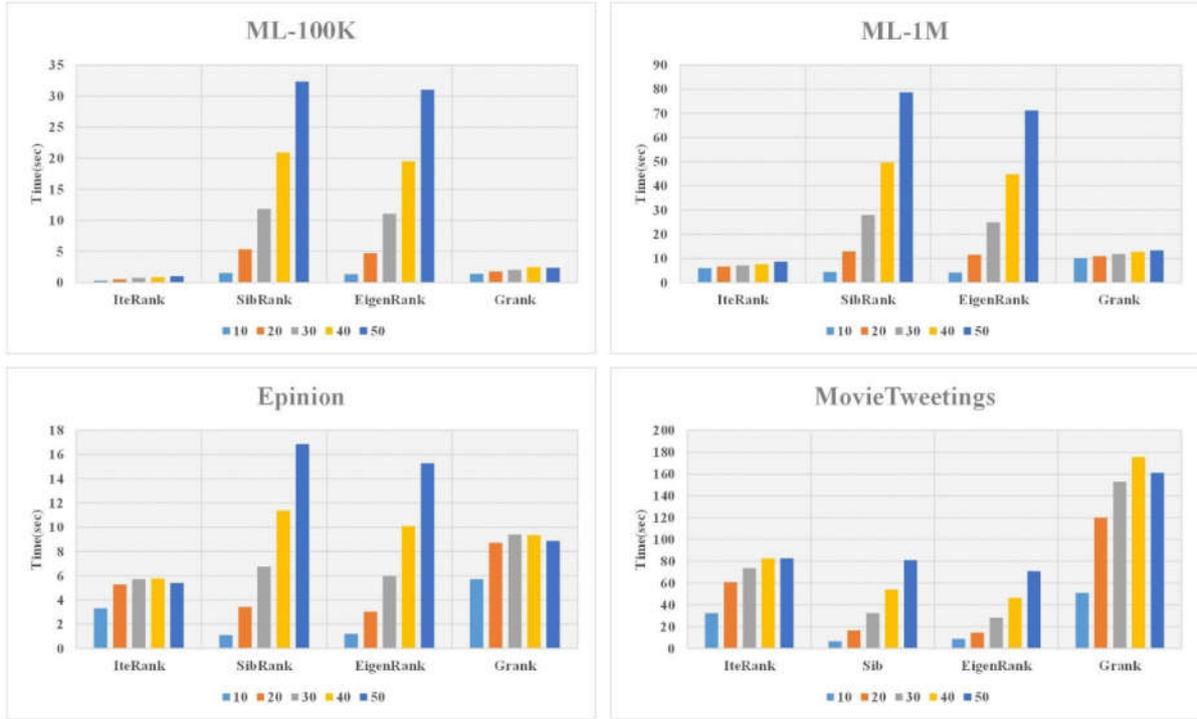

Fig.5. Comparison of NCR algorithms in terms of running time of recommendation process.

## 5. Discussion

As we mentioned before, this article presented a novel framework called IteRank, to address two main issues of traditional NCR algorithms: *Rare common preferences problem*, and, *low discrimination flaw*.

*Rare common preferences problem* of traditional NCR algorithms reduces their performance in two ways: first, they have to report zero similarity for pair of users with no common pairwise preferences. Second, even the non-zero similarity values are not reliable and differentiating enough as they are calculated based on a small number of common pairwise preferences for each pair of users. IteRank can address this issue as it assigns a concordance value to each possible pairwise preferences that is explored by any user. It then calculates users' similarities based on the whole set of pairwise preferences in a way that users who agree with highly concordant preferences are more similar to the target user. In other words, IteRank does not have to rely only on common pairwise comparisons to calculate users' similarities.

We report the fraction and levels of non-zero similarities for different algorithms in Table. 7. Due to limited space and similar trends of algorithms in different datasets, only the results corresponding to ML1M and Epinions have been reported to show the general concept. As shown in Table.7, IteRank has managed to predict non-zero similarity values between almost all pairs of users, regardless of whether they have any common pairwise comparisons or not.

It also can be seen that the calculated similarities by IteRank are more distinguishing as the number of different levels in calculated similarity values is much more than that of other algorithms. That clearly explains the superiority of IteRank over EigenRank as presented in Table 5. For instance, in Ml-1M when UPL=50, EigenRank shows performance of 69% in terms of NDCG@1, while, IteRank achieves performance of 78%.

Table 7. Numerical analysis of calculated similarities in IteRank, SibRank, and EigenRank.

| Dataset | UPL | Number of users | Fraction of non-zero similarity values | | | Levels of non-zero similarity values | | |
|---|---|---|---|---|---|---|---|---|
| | | | IteRank | Sib | EigenRank | IteRank | Sib | EigenRank |
| ML-1M | 20 | 5297 | 1 | 1 | 0.038 | 5297 | 5297 | 10.68 |
| | 30 | 4796 | 1 | 1 | 0.127 | 4796 | 4796 | 24.08 |
| | 40 | 4297 | 1 | 1 | 0.261 | 4297 | 4297 | 46.42 |
| | 50 | 3938 | 1 | 1 | 0.411 | 3938 | 3938 | 78.91 |
| Epinions | 20 | 618 | 0.935 | 0.935 | 0.017 | 527 | 385.3 | 4.23 |
| | 30 | 420 | 0.972 | 0.972 | 0.042 | 397 | 356 | 7.428 |
| | 40 | 315 | 0.984 | 0.984 | 0.074 | 309 | 289.6 | 11.18 |
| | 50 | 242 | 0.983 | 0.983 | 0.113 | 239 | 226.3 | 15.07 |

Another shortcoming of the traditional NCR framework is its *low discrimination flaw* that refers to the fact that they estimate zero concordance values for a large number of pairwise preferences that are not explored by the neighbors of the target user. IteRank tries to solve this problem by estimating and refining the concordance values in two phases. In the first phase, IteRank can estimate the concordance values for all pairwise preferences that are explored by any user. In this phase, all users that agree with a pairwise preference contribute to estimation of that preference's concordance to the target user, and the amount of their contribution is proportional to their similarity to the target user. Results show that using this approach IteRank has been able to significantly increase the fraction of non-zero estimated concordances (See Table 8).

However, there is still a large number of pairwise comparisons that are not explored by any user. Therefore, IteRank, cannot estimate their concordance values in its first phase. To resolve this issue, IteRank takes into account the overall desirability of items for estimating those concordance values. For instance, IteRank can estimate concordance values of preference *"A>B"* through considering *significance* of desirable representative of *A*, and undesirable representative of *B*. It is worth reminding that the significance values of the representatives are calculated using the concordance values of the explored preferences involving *A* or *B. (e.g. A>C and C>B")*. This phase allows IteRank to estimate non-zero concordance values for more than 99% of preferences, and has managed to discriminate among them well, as the number of different levels in calculated concordance values is much more than that of other algorithms. For instance, in Epinions dataset, EigenRank could estimate non-zero values for less than 0.001 of 25,000,000 possible pairwise preferences and has estimated concordance value of 1 or -1 for all of them, while IteRank has estimated non-zero concordance values for 100% preferences in 2,177,774 different levels of concordance. That is why IteRank significantly outperforms SibRank and EigenRank in sparse Epinions dataset in which there are a large number of non-explored pairwise preferences.

Table 8. Numerical analysis of concordance values in IteRank, SibRank, and EigenRank.

| Datasets | Fraction of non-zero concordance values | | | | Levels of concordance values | | | |
|---|---|---|---|---|---|---|---|---|
| | Eigen Rank | Sib Rank | IteRank-P1 | IteRank-P2 | Eigen Rank | Sib Rank | IteRank-P1 | IteRank-P2 |
| ML-1M | 0.001 | 0.002 | 0.032 | 1 | 11 | 304 | 106258 | 9972116 |
| | 0.004 | 0.004 | 0.055 | 1 | 7 | 656 | 237159 | 10722239 |
| | 0.006 | 0.006 | 0.079 | 1 | 11 | 1536 | 392782 | 11132049 |
| | 0.010 | 0.001 | 0.101 | 1 | 13 | 2306 | 557969 | 11235734 |
| Epinions | 0.000 | 0.000 | 0.002 | 1 | 3 | 2 | 2159 | 2177774 |
| | 0.000 | 0.001 | 0.003 | 1 | 18 | 18 | 2799 | 2371987 |
| | 0.000 | 0.002 | 0.004 | 1 | 27 | 40 | 3243 | 2221080 |
| | 0.002 | 0.003 | 0.005 | 1 | 64 | 100 | 3482 | 1820449 |

## 6. Conclusion

Here, we presented a novel NCR framework, called IteRank, that enables iterative refinement of similarities of users, concordance of pairwise preferences, and score of items. IteRank uses two novel network structures to improve estimation of pairwise concordance, and, consequently overall performance of the recommendation algorithms.

IteRank exploits a bipartite network UPNet in order to refine users' similarities and preferences concordance. It surfs all the user nodes of UPNet and takes into account their opinions based on their similarity to the target user; the random walker more frequently reaches the users with higher similarity, and, consequently, their opinions are more important in concordance estimation. This approach empowers IteRank to mainly cover the *Rare common preferences problem*. Also, IteRank uses another network structure, PRNet, for refining items' score and preferences' concordance. This structure enables IteRank to use overall scores of the items to refine the concordance values and to address the *low discrimination flaw*.

Experimental results showed that the key advantage of IteRank over former algorithms is its ability to estimate the concordance of pairwise preference based on the preferences of the similar users as well as the overall desirability of the items. Note that the traditional NCR algorithms, that efficiently capture the local tastes, showed a better performance in dense datasets while CofiRank, a well-known MFCR method, that learns the global desirability of items, achieved better results in sparse datasets.

There are several interesting directions to improve IteRank in the future works; it can be subject of further research to improve its accuracy and computational complexity through redesigning/punning the networks' structure or introducing new definitions for the concepts of user similarity and preference concordance.